\documentclass[a4paper,11pt]{article}
\pdfoutput=1 

\usepackage{jinstpub} 
\usepackage{tablefootnote}
\usepackage{multirow}
\usepackage[flushleft]{threeparttable}
\usepackage{hyperref}
\usepackage{lineno}

\title{\boldmath Improved bunching and longitudinal emittance control in an RFQ}

\author[a]{Robert A. Jameson}

\author[b,1]{Bruce Yee-Rendon,\note{Corresponding author.}}

\affiliation[a]{IAP, Goethe-Universitat Frankfurt, Frankfurt am Main D60438, Germany}

\affiliation[b]{J-PARC Center, Japan Atomic Energy Agency, 2-4, Shirakata, Tokai-mura,\\ 
Naka-gun, Ibaraki 319-1195, Japan}

\emailAdd{byee@post.j-parc.jp}

\abstract{A new application of vane modulation variation in a Radio Frequency Quadrupole (RFQ) cell has been applied that significantly improves beam bunching and longitudinal emittance control to achieve lower longitudinal rms emittance at the RFQ output. This procedure occurs in the individual cells, is independent of the overall design, and therefore is general, affording an extra parameter for beam manipulation. It can be applied besides the usual goals of vane modulation variation, e.g., to achieve higher acceleration efficiency. Examples of the cumulative effects on the overall design are provided to point out further exploration avenues for the designer.}


\begin{document}
\maketitle
\flushbottom

\section{Introduction}
\label{Sec:Intro}
The Radio Frequency Quadrupole (RFQ) operates at low energy, where control of the beam properties is critical to obtain the required beam quality in following particle accelerators~\cite{RFQ}. Achieving a small longitudinal rms output emittance is difficult because the initial bunching of the injected dc beam, initially completely emittance dominated, tends to fill the evolving bucket; the effective longitudinal rms emittance becomes large and cannot be reduced without an irreversible technique with beam loss.

The advanced application was developed in the updated RFQ equipartitioned (EP) design being developed by JAEA for an accelerator-driven subcritical system (JAEA-ADS)~\cite{ACC1}. The procedure is implemented at the cell level and, therefore, general. Examples of the cumulative effects on a JAEA-ADS and the International Fusion Materials Irradiation Facility (IFMIF) Conceptual Design Reference (CDR) EP RFQ~\cite{Bob1} overall designs are presented. The specifications of these two RFQs are reported in Table~\ref{tab:gral}. Other designs, e.g., a constant r0 RFQ (unpublished result), exhibit the same characteristics.

The design and simulation are carried out in the LINACS framework~\cite{Bob2}, with its open-source codes LINACSrfqDES and LINACSrfqSIM affording complete control of all RFQ design and simulation parameters, including beam-driven design, fully 3D simulation using precise quadrupolar symmetry with a rigorously connected Poisson solution for external and space charge fields, relatively short running time, and extensive analysis techniques.

The RFQ design is divided into three sections: radial matcher, shaper, and main RFQ, also called the accelerating section. The major difference between our design and the conventional proposed by Crandall et al.~\cite{RFQ2}, despite the reduction of the number of regions, is the flexibility to implement advanced parameter and beam-based space charge physics rules in the shaper and main RFQ. The default (standard) shaper design uses simple piece-wise linear rules for modulation and synchronous phase ($\phi_s$) to bring the injected beam into the equipartitioned (EP) condition~\cite{Bob3}  as soon as possible, at $\phi_s$ about –88$^\circ$, very near the pure bunching $\phi_s = -90^\circ$. This provides a smooth transition into the following acceleration section of the RFQ, avoiding problems with other shaper designs.  

\begin{table}[!htbp]
         \centering	
	\begin{threeparttable} 
		\caption{\label{tab:gral} RFQ specifications for the equipartitioned (EP) JAEA-ADS and IFMIF CDR models.}
		\begin{tabular}{|l|c|c|}
			\hline
			\textbf{Parameter}& \textbf{EP JAEA-ADS }& \textbf{IFMIF CDR (EP)} \\
			\hline
                            Particle    & Proton & Deuterons \\
			Beam current (mA ) & 20 & 130-140 \\
			Frequency (MHz) & 162 & 175  \\
			Initial/ Final energy (MeV) & 0.035/2.5 & 0.095/5  \\
			Duty factor  & CW  &CW \\
			Final $\varepsilon_{trans,rms}$ ($\pi$ mm mrad)& 0.21 & 0.27 \\
			Relevant specifications & High availability,  & Hands-on maintenance\\
			  &  EP= 1 and Kp= 1.2 &  EP, ratios vary 1.6 to 1.2, Kp=1.7 \\
			\hline
		\end{tabular}

	\end{threeparttable}
\end{table}

The EP condition establishes an internal energy balance relationship of the instantaneous state of the system in the time domain between the different degrees of freedom of the beam; therefore, there is no free energy to drive parametric resonances. The EP condition is represented as
\begin{equation}\label{eq:equip0}
     \frac{\varepsilon_{long, rms}}{\varepsilon_{trans,rms}}\frac{\sigma_{long}}{\sigma_{trans}}=1,
\end{equation}

where $\varepsilon_{long,rms}$ is longitudinal rms emittance, $\varepsilon_{trans, rms}$ is the transverse rms emittance, $\sigma_{long}$ is the longitudinal rms phase advance, and $\sigma_{trans}$ is the transverse rms phase advance. From the EP condition, Eq.~\ref{eq:equip0}, the corresponding ratios are   

\begin{equation}\label{eq:equip1}
     \frac{\varepsilon_{long,rms}}{\varepsilon_{trans,rms}}=\frac{\sigma_{trans}}{\sigma_{long}},
\end{equation}

EP provides efficient control of emittance growth in high-intensity hadron linacs~\cite{ACC1,Bob4} and in RFQs~\cite{Bob1,ACC2} . The RFQ shaper design above ensures that the beam space charge physics is under full control at least at one point in the RFQ, after which EP can be maintained or not. EP is not a mandatory choice for the accelerating section, but is a design trade-off. An alternative option is to design a matched trajectory that avoids major resonances and operates in regions where resonance interaction is small or quickly traverses resonant regions.  EP has advantages and disadvantages that need trade-offs and compromises for a particular project. 

\begin{figure}[!htb]
\centering
\includegraphics[keepaspectratio,scale=0.4]{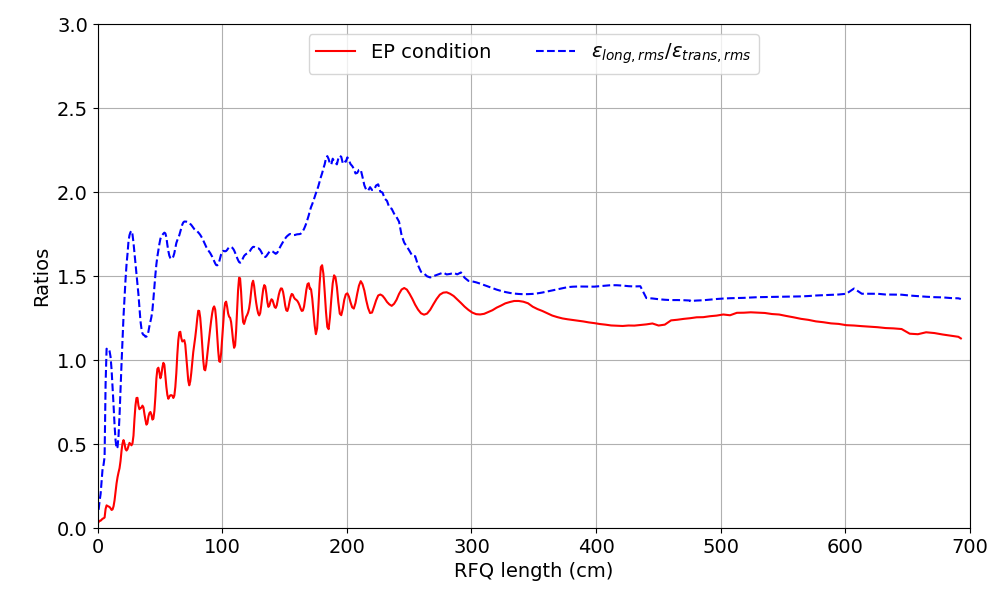}
\caption{The EP condition (solid red line) and emittances ratio (dashed blue line) for a JAEA-ADS design using pure 2-term longitudinal vane modulation. The design goals were to achieve a value of 1 for the EP and 1.25 for the emittance ratio.}
\label{fig:rfq_0}
\end{figure}

In our case, we adopted the EP condition because it ensures low emittances, tight beams, and low loss.  The following discussion is for the JAEA-ADS specification calling for the minimization of RFQ rms longitudinal output emittance. 
For the JAEA-ADS, the emittances ratio, given by Eq.~\ref{eq:equip1}, was set to 1.25.  The usual vane modulation procedures were unable to meet the requirements, as presented in figure~\ref{fig:rfq_0}. Thus, an advanced approach was implemented.

\begin{figure}[!htb]
\centering
\includegraphics[keepaspectratio,scale=0.4]{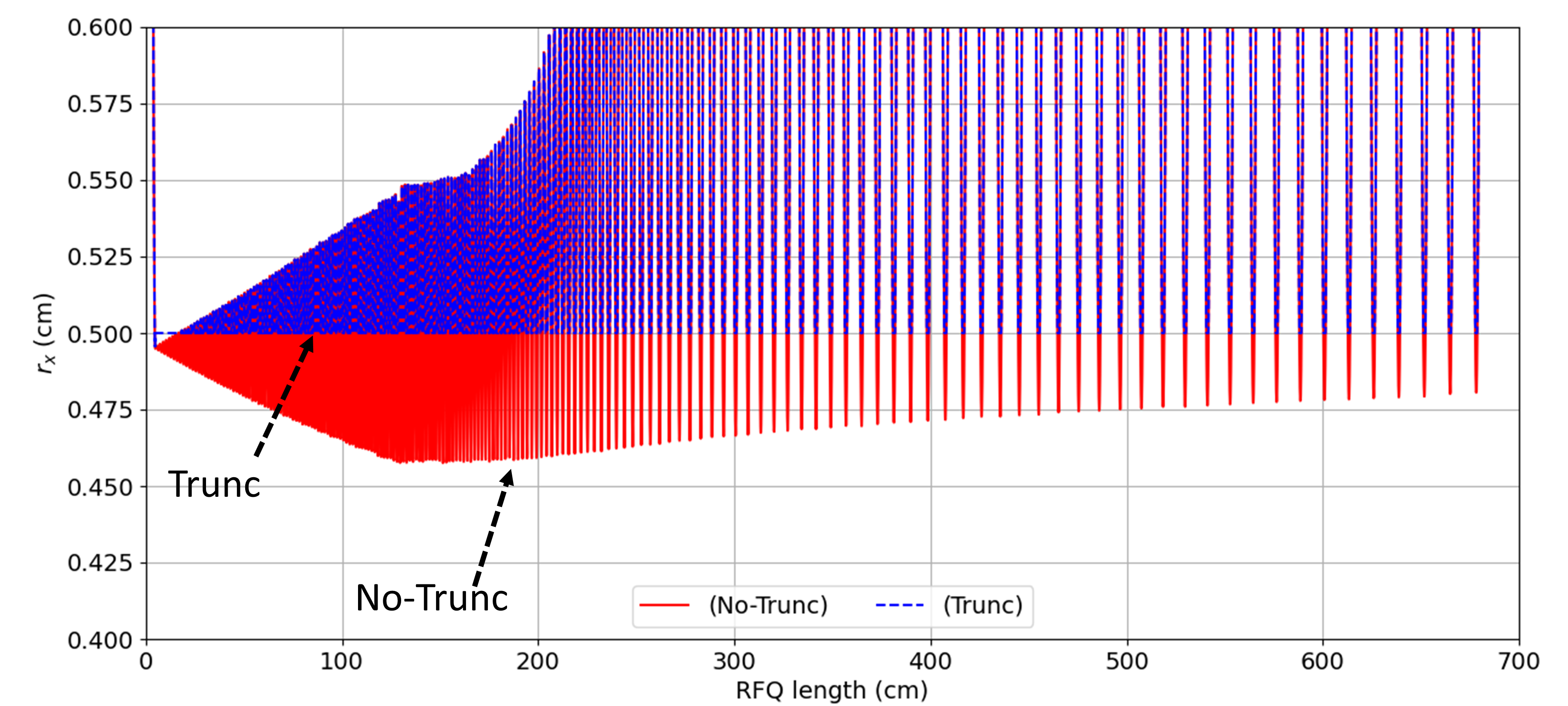}
\caption{The horizontal 2-term vane radii for the No-Truncated (No- Trunc, solid red line) and Truncated (Trunc, dashed blue line). The underlying No-Truncated design has a minimum radius of 0.46 cm at the end of the No-Truncated shaper, then increases. }
\label{fig:vane_0}
\end{figure}

The initial longitudinal vane modulation in the example here is 2-term, from the standard multipole representation of the external potential, but the choice is arbitrary. After the shaper, the aperture is allowed to increase slightly (the inner vane tip profile) out to a radius of 0.6 cm, as shown by the solid red line in figure~\ref{fig:vane_0}.  Initial designs for the JAEA-ADS EP RFQ used a very long, gradual shaper section ending at 129 cm, cell 161, which helped reduce the output $\varepsilon_{long,rms}$ but could not meet the specifications. The output $\varepsilon_{long,rms}$ and EP condition requirements are very severe and not achievable using the standard shaper, which is also used in the IFMIF CDR EP RFQ.

The vanes were then cut off, in the simulation, the so-called truncated vane, at a radius of 0.5 cm, display by the blue dashed line in figure~\ref{fig:vane_0}.

\section{The \emph{truncated shaper and following section} – standard rms smooth approximation diagnostics}
\label{Sec:trunc}

\begin{figure}[!htb]
\centering
\includegraphics[keepaspectratio,scale=0.9]{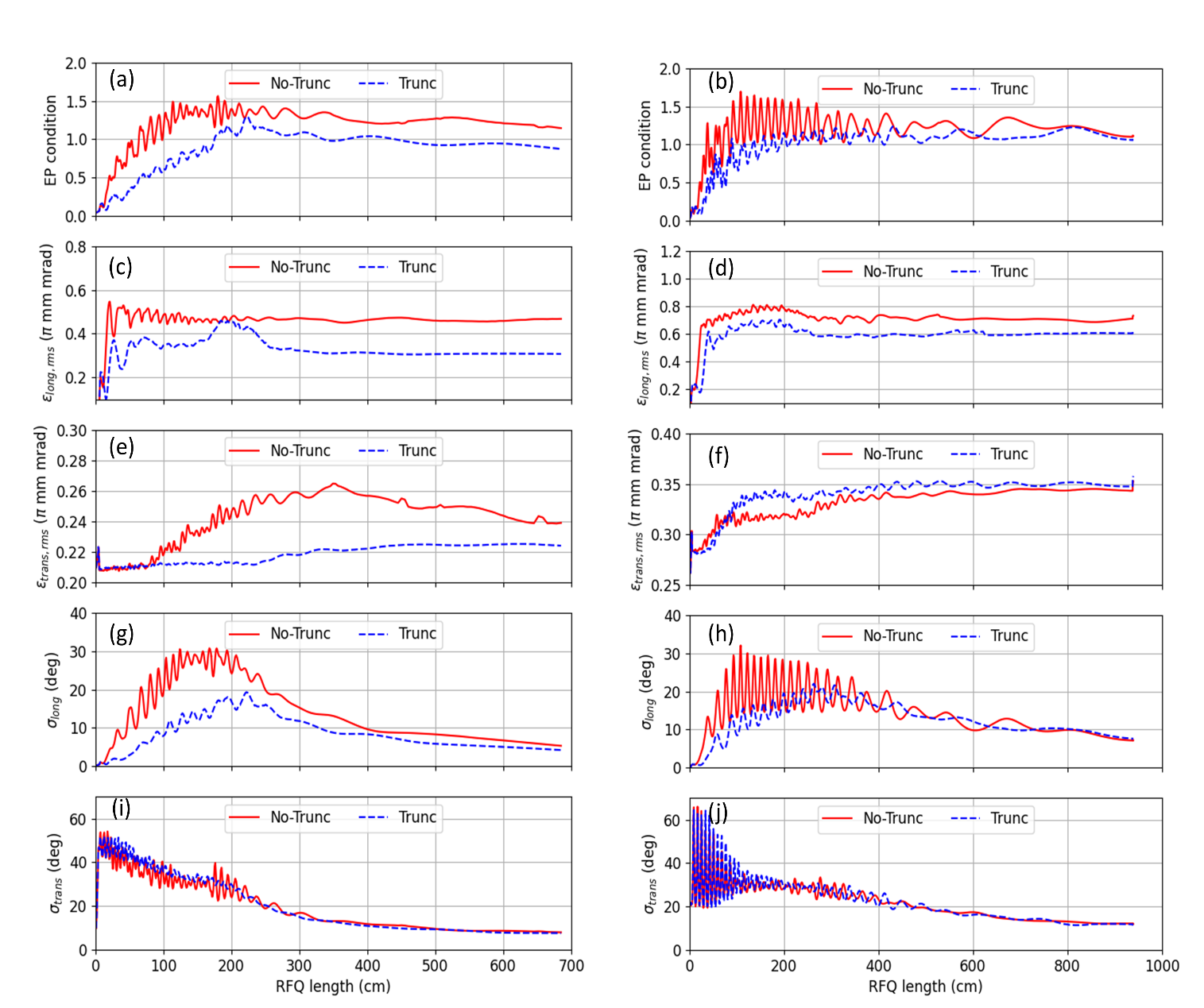}
\caption{Design comparison between Truncated and No-truncated vanes for the JAEA-ADS EP RFQ (left) and the IFMIF CDR EP RFQ (right). (a) and (b) plot the EP condition, (c) and (d) are the longitudinal normalized rms emittance, (e) and (f) are the transverse normalized rms emittance, (g) and (h) are the longitudinal phase advance, and (i) and (j) are the transverse phase advance.}
\label{fig:comp_0}
\end{figure}

Figure~\ref{fig:comp_0} reveals a significant reduction in the output longitudinal rms emittance in subplots (c) and (d) and an improved EP condition in subplots (a) and (b). The right side of figure~\ref{fig:comp_0} shows the performance of the IFMIF CDR RFQ with a corresponding truncation. Table~\ref{tab:gral2} summarizes the transmitted (Xmsn) and accelerated (Accel) particles and the normalized rms emittances for the two RFQ models with and without truncated vanes.

Most of the improved performance occurred in the shaper region. However, the truncation beyond the shaper also plays an important role in forming and maintaining the improvements in longitudinal emittance and EP performance, as was found compared to cases with truncation of only the shaper. Truncation after the shaper will depend on the design and will have to be handled on a case-by-case basis.

With the truncated vane profile, the standard rms smooth approximation diagnostics revealed that the $\sigma_{long}$ had become linear in the shaper; this appears general and is clarified below. The truncation radius resulting in linear $\sigma_{long}$ was slightly larger than the radius at the beginning of the shaper, i.e., the end of the radial matching section; this may also depend on the underlying shaper modulation details. It was necessary to move to the broader view of the instantaneous system state to reveal the process that occurred.

\begin{table}[!htbp]
         \centering	
	\begin{threeparttable} 
		\caption{\label{tab:gral2} The percentage of transmitted (Xmsn) and accelerated (Accel) particles, and the output rms longitudinal ($\varepsilon_{trans,rms}$) and transverse ($\varepsilon_{trans,rms}$)  for the JAEA-ADS EP and IFMIF CDR RFQ models.}
		\begin{tabular}{|l|c|c|c|c|}
			\hline
			\textbf{Parameter}&  \multicolumn{2}{c|}{\textbf{EP JAEA-ADS }}& \multicolumn{2}{c|}{ \textbf{IFMIF CDR (EP)} }\\
			\cline{2-5}
			\textbf{ }& \textbf{Trunc} & \textbf{Not-Trunc }& \textbf{Trunc}& \textbf{Not-Trunc} \\
			\hline
                            Xmsn (\%) & 99.71 & 97.71 & 99.52 & 99.73\\
                            Accel (\%) & 98.87 & 97.70 & 98.67 & 99.51\\  
			$\varepsilon_{long,rms}$ ($\pi$ mm mrad)& 0.32 & 0.47 & 0.61 & 0.70 \\

			$\varepsilon_{trans,rms}$ ($\pi$ mm mrad)& 0.22 & 0.24 & 0.35 & 0.35 \\
			\hline
		\end{tabular}

	\end{threeparttable}
\end{table}

\section{Time-domain diagnostics }
\label{Sec:TDD}

\begin{figure}[!htb]
\centering
\includegraphics[keepaspectratio,scale=0.9]{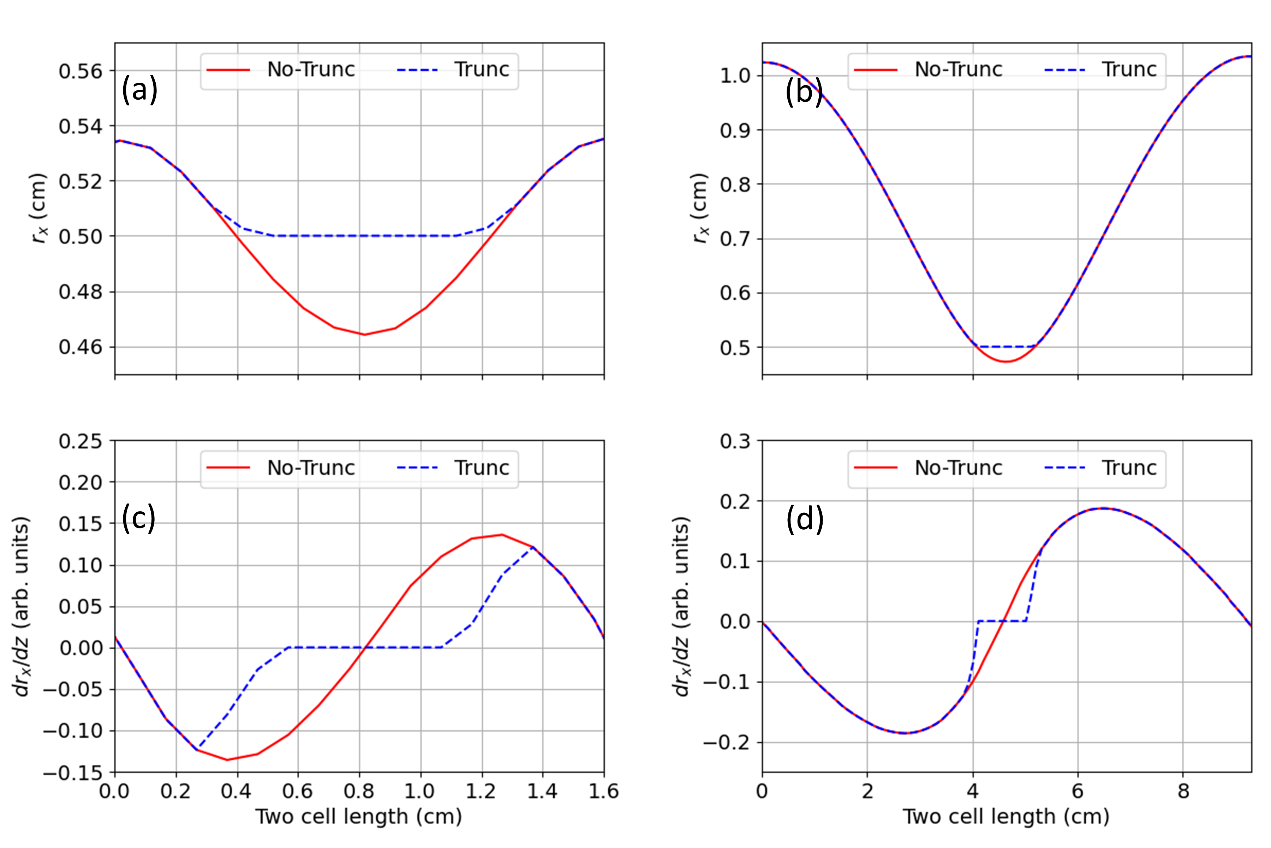}
\caption{Horizontal vane radii, (a) and (b), and their corresponding resulting electric field shape ($dr/dz$), (c) and (d), in two cells for the No-Truncated (solid red line) and Truncated (dashed blue line) are plotted at two different locations of the RFQ: 100 cm (last part of the shaper, left) and 400 cm (middle of the acceleration section, right).}
\label{fig:Vanes_2}
\end{figure}

The profile is truncated, very significantly in the shaper, and in the acceleration part, as shown in figure~\ref{fig:Vanes_2} top and the corresponding electric field on the bottom.

\begin{figure}[!htb]
\centering
\includegraphics[keepaspectratio,scale=0.4]{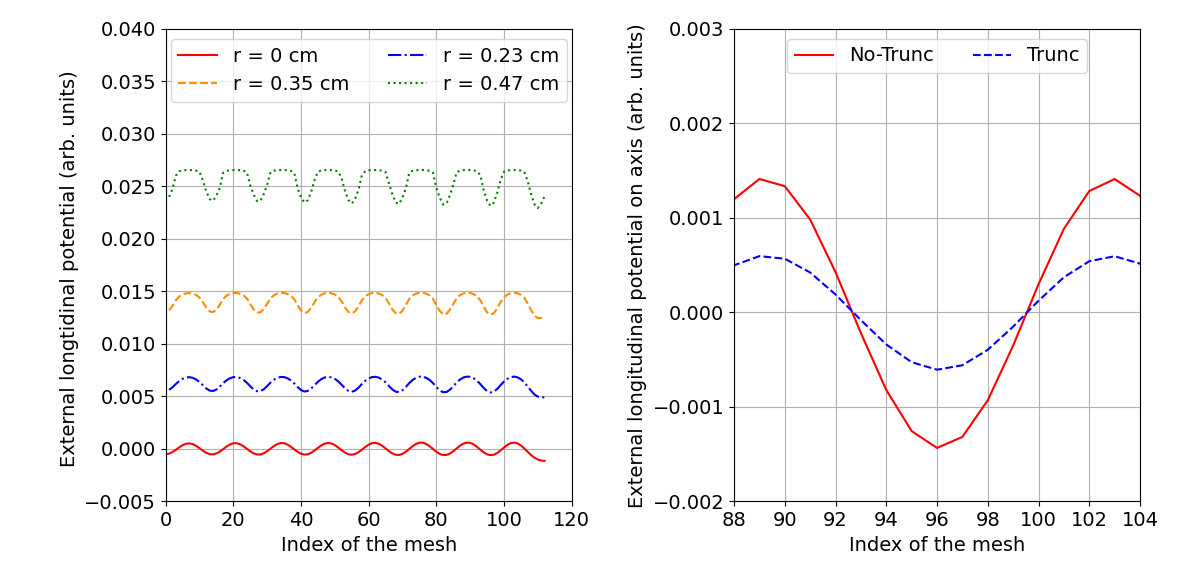}
\caption{On the left, external longitudinal potential at different transverse radii in the cell 102 (near the end of the shaper); the field near the vane has a truncated shape ($r$ equal to 0.47 cm). The right plot is the longitudinal potential on axis for the No-Truncated and Truncated cases.}
\label{fig:pot}
\end{figure}

The explanation is seen from an expansion of figure~\ref{fig:Vanes_2} to figure~\ref{fig:pot}. Figure~\ref{fig:pot} (left) displays the external longitudinal potential at different transverse radii in the cell 102 (near the end of the shaper), from which the resulting fields will interact with the beam in the time domain. \cite{Wan,Trap1,Trap2,Trap3}\footnote{The basics of RF acceleration, longitudinal particle dynamics, and application to the RFQ are well-covered in Ref.~\cite{Wan}, and extension to vane profiles with flat sections (trapezoidal) in references~\cite{Trap1,Trap2,Trap3}; this background is assumed.  The usual emphasis is on the vane tip profile.  The resulting on-axis longitudinal field is found by the Poisson solver.}  Figure~\ref{fig:pot}  (right) shows the total field of the horizontal and vertical vanes on the axis, i.e., transverse radius equal to zero. There is a change, but what is happening is not so evident from this.

Here, the main interest was to influence the bunching such that the longitudinal emittance is kept as low as possible, as shown in figures~\ref{fig:comp_0} (c) and (d). Thus, a region of zero longitudinal fields was introduced by truncation of the inner vane radius, figure~\ref{fig:Vanes_2} bottom, leaving the modulation in the remaining part of the cell at the full depth, figure~\ref{fig:pot} left. Here of a 2-term vane modulation, a sinusoidal would be equivalent, and it will be shown below that trapezoidal is equivalent.

There is a longitudinal external field and acceleration from a vane only when the vane profile is changing; in this case, driving the bunching around $\phi_s$ that is approximate –90$^\circ$ in the shaper. Here, the outer part of the profile is not truncated, and bunching is driven, with a larger transit time factor over the cell. The inner radii are truncated, and there is no longitudinal action over the truncated region; the $\pm$ swing is set to zero over that region. In the truncated zone, the beam will continue to bunch with the velocities it has picked up outside the truncated region. Thus, the whole bucket does not get filled.

In this example, the underlying vane design was unchanged; the truncation applied to the vane profile in the Poisson simulation. The design modulation beyond the truncation point to the full depth, approximately modulation times aperture, was preserved.  The amount of truncation was optimized for lowest output rms longitudinal emittance, which also produced the most linear longitudinal phase advance in the shaper.

\begin{figure}[!htb]
\centering
\includegraphics[keepaspectratio,scale=0.9]{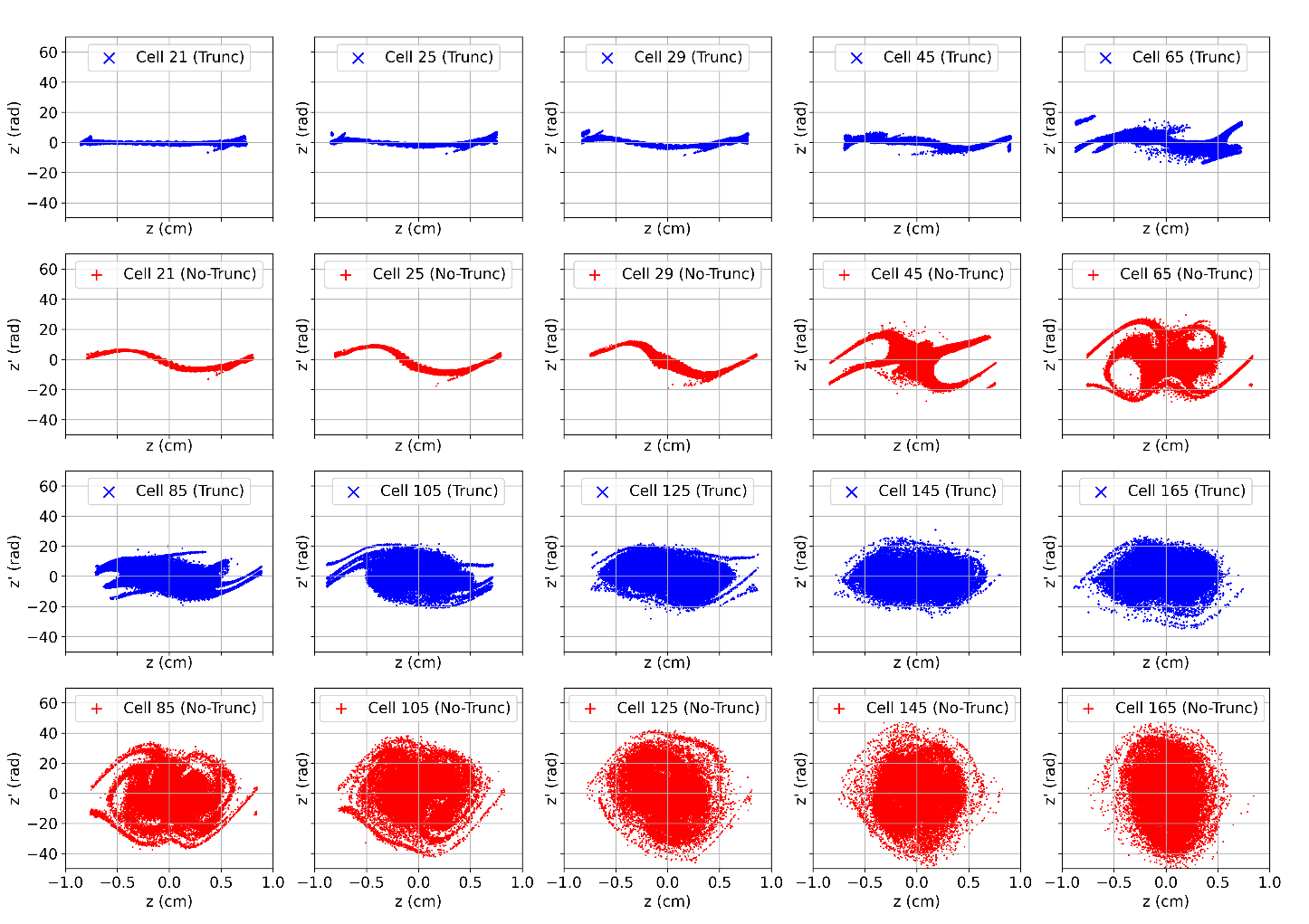}
\caption{Longitudinal phase distribution at different locations along the shaper section for Truncated vane (blue x) and No-Truncated (red plus).}
\label{fig:space}
\end{figure}

\begin{figure}[!htb]
\centering
\includegraphics[keepaspectratio,scale=0.55]{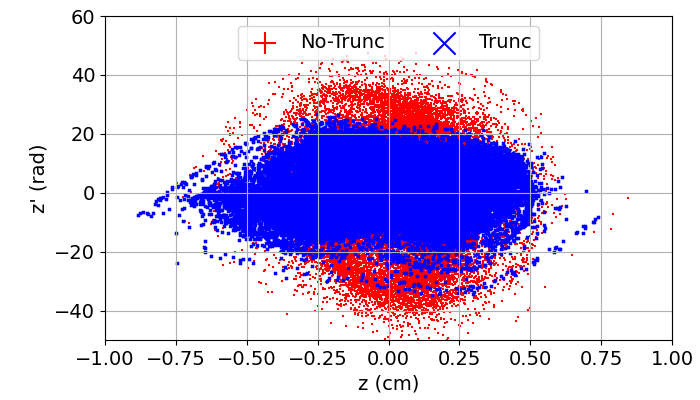}
\caption{Longitudinal phase space comparison between Truncated vane (blue x) and No-Truncated (red plus) at the end of the shaper, cell 165.}
\label{fig:long}
\end{figure}

\begin{figure}[!htb]
\centering
\includegraphics[keepaspectratio,scale=0.65]{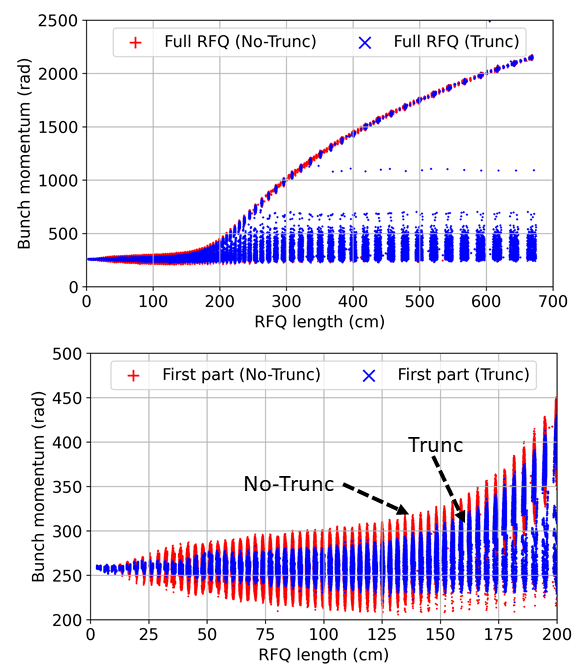}
\caption{Longitudinal phase space comparison between Truncated vane (blue x) and No-Truncated (red plus). The top plot shows the full RFQ, and the bottom one expanded for the shaper section and somewhat beyond to 200 cm.}
\label{fig:long_2}
\end{figure}

Figures~\ref{fig:space}-\ref{fig:long_2} are the $z-z’$ phase-space plots, showing the bunching and acceleration processes of a distribution of 1~$\times~ 10^{5}$ macroparticles.

Figure~\ref{fig:long} and Table~\ref{tab:gral2} provide information on the evolution of the particle distribution. Without truncation, there are more radial losses, and the transmitted particles and the particles inside the energy band, which in this case was $\pm$50\% of the synchronous energy, were the same. With truncation, there was less transmission loss, particles that were not in the selected energy range were transmitted to the end of the RFQ, and the space-charge effect during the bunching process resulted in a tighter bunch with less energy spread. 
 
\section{Extension to trapezoidal case – truncation of both inner and outer radii}
\label{Sec:Trape}

A different strategy using zero-longitudinal-field regions inside the cell - trapezoidal vane modulation - has long been used to enhance the RFQ acceleration efficiency at higher modulations $\geq$~$\approx$2.

\begin{figure}[!htb]
\centering
\includegraphics[keepaspectratio,scale=0.4]{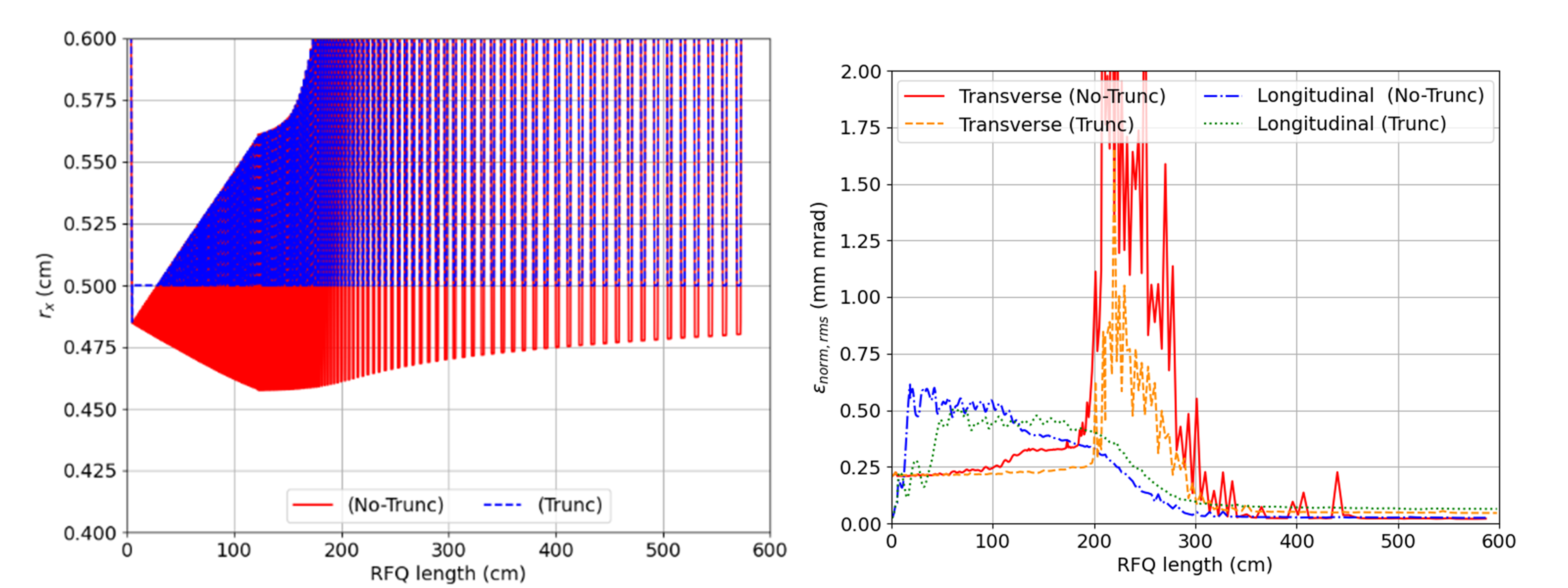}
\caption{Characteristics of trapezoidal vane modulation. Vane profiles (left) and emittance performance (right) comparison.  There are heavy beam losses from 230 cm.}
\label{fig:trap}
\end{figure}

Comparison to trapezoidal modulation was implemented by changing only the vane modulation type, with all other design and simulation parameters left the same as for the 2-term modulation case.  Figure~\ref{fig:trap} (left) compares trapezoidal (40\%/cell flat regions) vane profiles without and with truncation.  For this discussion, the shaper section is of primary interest.

Figure~\ref{fig:trap} (right) compares the longitudinal emittance behavior. The same behavior in the shaper as for the 2-term vanes was expected because the detailed shape of the longitudinal vane modulation is a minor change. Beyond the shaper, additional redesign in the main acceleration section is indicated.

\section{On the R\&D of advanced processes using LINACS  }
\label{Sec:RD}
A \emph{design RFQ} is established, using a choice for the vane transverse and longitudinal geometry (currently sinusoidal, trapezoidal, and 2-term longitudinal modulation), circular vane tip, and tapered or straight blade vane body; plus rules, defined in terms of $\beta$ or $z$, for aperture, vane voltage, modulation, and synchronous phase throughout the RFQ as the design progresses.  Or, one or more of these parameters are left free to simultaneously solve, as the design proceeds, the matching transverse and longitudinal equations for the rms radii of the beam and one or more equations involving some function(s) of the accessible rms quantities, such as requiring equipartition.

A table of pre-computed 8-term multipole coefficients (obtained from an accurate Poisson simulation of the vane's detailed geometry) and transit-time factor, peak field, and intervane capacitance is then accessed.  The table is parameterized by Rho (radius of curvature of vane tip), Ls = (cell length)/(average radius of vane tip from axis = r0 = r0rfq), and the modulation = emrfq.  Local interpolation produces the generated design cell table that tabulates all cell parameters at the end of each cell. The 8-term potential approximation is not appropriate for simulation, but it is well applied in the design procedure and allows a good agreement with the design simulation.

This \emph{design RFQ} cell end table is then used in the simulation program to generate the spatial and external Poisson meshes, with detailed design vane geometry, cell lengths, aperture, and modulation. The results of the Poisson simulation accurately describe the physical process that occurs.  If beam-based, \emph{inside-out} requirements have been imposed on the design, the simulated results may or may not agree well with the design, depending on the actual physical practicality of the design. Generally, it is desired to have the design and simulation agree well, so if a disagreement occurs, the design process is iterated. If an appropriate rms smooth approximation design is found, experience indicates that the experimental and simulation measurements in the constructed RFQ will also agree well.

The simulation is free to introduce off-design condition; in which case the results can be expected to deviate from the design condition results.  

Here we have developed a control for bunching and longitudinal emittance, quantities inaccessible to the design equations. Imposing a truncation on an inherently smooth vane modulation (2-term or sinusoidal) did not interfere with the underlying design modulation, produced desirable simulation results, and allowed optimization of the truncation. But if the underlying modulation had already used truncation (such as trapezoidal modulation) to control something else, such as the acceleration efficiency, then implementing additional truncation in the simulation to influence an additional aspect might be conflicting and perhaps require a compromise, which could require a complicated optimization process to find. 

The point is that for good agreement between the design and simulation, the underlying design tables must be derived separately for any particular form of vane modulation, i.e., for any specific form of intra-period manipulation.

    This study and subsequent research applying intra-period detail manipulation require open-source correlated design and simulation tools capable of precise physics. In addition, ease of modification, use, and run-time, such as LINACS, must be reasonable. In addition, LINACS can also compute new multipole tables for a new variation. Although this is time-consuming and to achieve high precision, another program may be more convenient~\footnote{LINACS tables for 2-term, sinusoidal and trapezoidal longitudinal vane modulation are actually computed to high precision using TOSCA.}.

\section{Conclusions}
\label{Sec:CO}

An improved vane longitudinal modulation profile technique for longitudinal emittance control is presented.  This advanced approach applies manipulation of the RFQ vane modulation profiles to minimize longitudinal rms emittance. It is implemented at the cell level, and therefore general, without change to the overall rms design, e.g., while maintaining the requirements to reach equipartition at the end of the shaper and maintain equipartition in the acceleration section to the end of the RFQ. The technique can be adopted by any RFQ, of course, also with attention to the details of the case at hand.  

The key is to avoid most of the deleterious longitudinal field variations, particularly in the bunching process, by truncating the vane profile so that no longitudinal electric field is produced in that region.

This work also points out that the limited \emph{rms/smooth-approximation/steady-state} approach gives useful analysis insights.  However, it has to be expanded to the full time-domain, instantaneous-state view, where clearly much is going on within the period.  All the rms observations are also apparent in the time domain view, but smoothing all that away, i.e., averaging over a distance of interest and discarding detailed motion within that distance, lost the ability to explain what was happening. Here, the phase-space details provide helpful information.

\section*{Acknowledgments}
Bruce would like to acknowledge the financial support of the Ministry of Education, Culture, Sports, Science, and Technology (MEXT) through the Subsidy for Research and Development on Nuclear Transmutation Technology in the Japan Atomic Energy Agency (JAEA).  

\end{document}